\documentclass[12pt,preprint]{emulateapj}
\usepackage{natbib}
\slugcomment{Accepted for publication in {\it Ap. J. Letters GALEX Special Issue}}

\shorttitle{Evolution of the Far-Ultraviolet Luminosity Density}
\shortauthors{Schiminovich et al.}

\begin{document}


\title{The GALEX-VVDS Measurement of the Evolution of the Far-Ultraviolet Luminosity\\
 Density and the Cosmic Star Formation Rate}

\author{
D. Schiminovich\altaffilmark{1,2}
O. Ilbert\altaffilmark{3},      
S. Arnouts \altaffilmark{3}, 
B. Milliard\altaffilmark{3},
L. Tresse\altaffilmark{3},      
O. Le F\`evre\altaffilmark{3}, 
M. Treyer \altaffilmark{2,3},   
T. K. Wyder\altaffilmark{2},
T. Budav\'ari\altaffilmark{4},    
E. Zucca  \altaffilmark{5},
G. Zamorani  \altaffilmark{5}, 
D. C. Martin\altaffilmark{2}, 
C. Adami \altaffilmark{3}, 
M. Arnaboldi \altaffilmark{6}, 
S. Bardelli \altaffilmark{5},
T. Barlow\altaffilmark{2}, 
L. Bianchi\altaffilmark{4},     
M. Bolzonella \altaffilmark{7},
D. Bottini \altaffilmark{8}, 
Y.-I. Byun\altaffilmark{9}, 
A. Cappi \altaffilmark{5}, 
T. Contini \altaffilmark{10},
S. Charlot \altaffilmark{11,12}, 
J. Donas \altaffilmark{3}, 
K. Forster\altaffilmark{2},     
S. Foucaud \altaffilmark{13},  
P. Franzetti \altaffilmark{13},
P. G. Friedman\altaffilmark{2}, 
B. Garilli \altaffilmark{13}, 
I. Gavignaud \altaffilmark{10}, 
L. Guzzo \altaffilmark{14}, 
T. M. Heckman\altaffilmark{4},  
C. Hoopes\altaffilmark{4},  
A. Iovino \altaffilmark{14}, 
P. Jelinsky\altaffilmark{15}, 
V. Le Brun \altaffilmark{3},
Y.-W. Lee\altaffilmark{9},      
D. Maccagni \altaffilmark{13}, 
B. F. Madore\altaffilmark{16}, 
R. Malina\altaffilmark{3},     
B. Marano \altaffilmark{7}, 
C. Marinoni \altaffilmark{3},
H.J. McCracken \altaffilmark{12}, 
A. Mazure \altaffilmark{3},
B. Meneux \altaffilmark{3},
P. Morrissey\altaffilmark{3},   
S. Neff\altaffilmark{17}, 
S. Paltani \altaffilmark{3},
R. Pell\`o \altaffilmark{10}, 
J.P. Picat \altaffilmark{10},  
A. Pollo \altaffilmark{14}, 
L. Pozzetti \altaffilmark{5},
M. Radovich \altaffilmark{6}, 
R. M. Rich\altaffilmark{18},    
R. Scaramella \altaffilmark{8}, 
M. Scodeggio \altaffilmark{13}, 
M. Seibert \altaffilmark{3}, 
O. Siegmund\altaffilmark{15}, 
T. Small\altaffilmark{2},       
A. S. Szalay\altaffilmark{4}, 
G. Vettolani \altaffilmark{19},
B. Welsh\altaffilmark{15},
C. K. Xu  \altaffilmark{20}, 
A. Zanichelli  \altaffilmark{19}
}

\altaffiltext{1}{Department of Astronomy, Columbia University, MC2457,
550 W. 120 St. NY,NY 10027}
\altaffiltext{2}{California Institute of Technology, MC 405-47, 1200 East
California Boulevard, Pasadena, CA 91125}
\altaffiltext{3}{Laboratoire d'Astrophysique de Marseille, BP 8, Traverse
du Siphon, 13376 Marseille Cedex 12, France}
\altaffiltext{4}{Department of Physics and Astronomy, The Johns Hopkins
University, Homewood Campus, Baltimore, MD 21218}
\altaffiltext{5}{Osservatorio Astronomico di Bologna, via Ranzani, 40127 Bologna,
 Italy }
\altaffiltext{6}{Osservatorio Astronomico di Capodimonte, via Moiariello 16,
 80131 Napoli, Italy }
\altaffiltext{7}{Universit\`a di Bologna, Departimento di Astronomia,
 via Ranzani 1, 40127 Bologna, Italy}
\altaffiltext{8}{IASF-INAF, Milano, Italy }
\altaffiltext{9}{Center for Space Astrophysics, Yonsei University, Seoul
120-749, Korea}
\altaffiltext{10}{Laboratoire d'Astrophysique - Observatoire Midi-Pyr\'en\'ees,
 Toulouse, France }
\altaffiltext{11}{Max Planck Institut fur Astrophysik, 85741 Garching, Germany}
\altaffiltext{12}{Institut d'Astrophysique de Paris, UMR 7095, 98 bis bvd Arago,
 75014 Paris, France }
\altaffiltext{13}{Laboratoire d'Astrophysique - Observatoire Midi-Pyr\'en\'ees,
 Toulouse, France }
\altaffiltext{14}{Osservatorio Astronomico di Brera, via Brera, Milan, Italy}
\altaffiltext{15}{Space Sciences Laboratory, University of California at
Berkeley, 601 Campbell Hall, Berkeley, CA 94720}
\altaffiltext{16}{Observatories of the Carnegie Institution of Washington,
813 Santa Barbara St., Pasadena, CA 91101}
\altaffiltext{17}{Laboratory for Astronomy and Solar Physics, NASA Goddard
Space Flight Center, Greenbelt, MD 20771}
\altaffiltext{18}{Department of Physics and Astronomy, University of
California, Los Angeles, CA 90095}
\altaffiltext{19}{Istituto di Radio-Astronomia-CNR, Bologna, Italy }
\altaffiltext{20}{NASA/IPAC, California Institute
of Technology, Mail Code 100-22, 770 S. Wilson Ave., Pasadena, CA 91125}
%
%
%\altaffiltext{19}{Observatoire de Paris, LERMA, UMR 8112, 61 Av. de l'observatoire,
% 75014 Paris, France }
%


\begin{abstract}

In a companion paper (Arnouts et al. 2004) we presented new measurements of the galaxy luminosity
function at 1500\AA	~out to z$\sim$1 using GALEX-VVDS observations (1039 galaxies with NUV$\leq$24.5 and z$>$0.2) and at higher z using existing data sets.  In this paper we use the same sample to study evolution of the FUV luminosity density $\rho_{1500}$.  We detect evolution consistent with a (1+z)$^{2.5\pm0.7}$ rise to z$\sim$1 and (1+z)$^{0.5\pm0.4}$ for z$>$1.  The luminosity density from the most UV-luminous galaxies (UVLG) is undergoing dramatic evolution ($\times 30$) between 0$<$z$<$1.  UVLGs are responsible for a significant fraction ($>$25\%) of the total FUV luminosity density at z$\sim$1.  We measure dust attenuation and star formation rates of our sample galaxies and determine the star formation rate density ($\dot{\rho_\star}$) as a function of redshift, both uncorrected and corrected for dust.   We find good agreement with other measures of $\dot{\rho_\star}$ in the rest ultraviolet and H$\alpha$ given the still significant uncertainties in the attenuation correction.

\end{abstract}


%% subject header.  Objects should be in the appropriate "individual"
%% headers (e.g. quasars: individual, stars: individual, etc.) with the
%% additional provision that the total number of headers, including each
%% individual object, not exceed six.  The \objectname{} macro, and its
%% alias \object{}, is used to mark each object.  The macro takes the object
%% name as its primary argument.  This name will appear in the paper
%% and serve as the link's anchor in the electronic edition if the name
%% is recognized by the data centers.  The macro also takes an optional
%% argument in parentheses in cases where the data center identification
%% differs from what is to be printed in the paper.



\section{Introduction}

The rest-frame far-ultraviolet (FUV; 1500\AA)  luminosity has been used to determine the star formation rate (SFR) of stellar populations over the complete range of redshifts for which galaxies have been observed.   The utility and limitations of the integrated measures---the FUV luminosity function ($\phi_{FUV}$) and luminosity density ($\rho_{1500}$)---and their relation to the star formation history of the universe has been extensively discussed and reviewed (e.g. Madau, Pozzetti, \& Dickinson 1998, Hopkins 2004).  A principal goal of the Galaxy Evolution Explorer (GALEX) mission (Martin et al. 2004) is to perform deep wide-angle surveys to obtain an accurate measurement of the evolution of the FUV luminosity density over the range $0<z<1$ and beyond.  In this letter we present results from a small pilot study performed in the 2h VIRMOS-VLT Deep Survey field using measurements from 1039 galaxies.

GALEX data will allow us to determine how the rest-UV can best  be used to study the detailed properties of galaxies (e.g. dust, metallicity, star formation history).  Since this is work in progress, here we will instead use existing  methods to determine the intrinsic luminosity of galaxies in the FUV (Meurer, Heckman, \& Calzetti 1999, hereafter MHC99, for dust corrections) and the SFR that this luminosity implies (Kennicutt 1998 for SFR conversion).  This simple analysis yields  some quick answers---we will discuss how
this work will be expanded and developed in the near future.

Throughout this paper we adopt the flat-lambda cosmology ($\Omega_M = 0.3, \Omega_\Lambda = 0.7$) with $H_0 = 70$ km s$^{-1}$ Mpc$^{-1}$.

\section{Data}

GALEX observations of the VVDS 0226-04 field (02h26m00s ~ -04$^\circ30\arcmin00\arcsec$, J2000) were performed in October-November 2004 as part of the GALEX Deep Imaging Survey.  Further details of these observations, the subsequent match to VVDS spectroscopy and photometry and the calculation of the LF can be found in the companion letter Arnouts et al. (2004) (hereafter Paper I) and references therein.  Paper I also describes the derivation of the $\phi_{FUV}$ at z=2.0 and 2.9 using an HDF sample from Arnouts et al. (2002).  For comparison we also use the local $\phi_{FUV}$ (Wyder et al. 2004) and the z$\sim$3 Lyman Break Galaxy (LBG) $\phi_{FUV}$ (Steidel et al. 1999)

\begin{deluxetable}{clll}
\scriptsize
\tablecaption{FUV 1500\AA\ Luminosity Density}
\tablehead{
        \colhead{$<z>$} &
        \colhead{log $\rho_{1500}^a$} &
        \colhead{log $\rho_{1500}^a$} &
        \colhead{log $\rho_{1500}^a$} \\
        \colhead{$$} &
        \colhead{$total$} &
        \colhead{$V_{max}$} &
        \colhead{$L>L_{min}^b$} \\
}
\startdata
{0.055$^{+0.045}_{-0.055}$}$^c$    &       25.54$^{+0.09}_{-0 .02}$       &      &     23.97   \\
\hline
0.3$\pm0.1$     &           25.86$^{+0.05}_{-0.05} $       &      25.86$^{+0.05}_{-0.05}$  &     24.67$^{+0.17}_{-0.18}$  \\
0.5$\pm0.1$     &           25.97$^{+0.15}_{-0.08}$        &     25.76$^{+0.05}_{-0.05}$   &     25.20$^{+0.09}_{-0.09}$ \\
0.7$\pm0.1$     &           26.16$^{+0.31}_{-0.13}$       &     25.91$^{+0.05}_{-0.05} $  &     25.48$^{+0.05}_{-0.05}$ \\
1.0$\pm0.2$     &      26.11$^{+0.31}_{-0.13}$$^e$      &      25.69$^{+0.05}_{-0.05}$   &     25.51$^{+0.05}_{-0.05}$ \\
\hline
2.0$\pm{0.5}$     &      26.45$^{+0.25}_{-0.09}$      &      26.30$^{+0.04}_{-0.04}$   &     26.03$^{+0.13}_{-0.12}$  \\
2.9$\pm{0.5}$     &      26.52$^{+0.17}_{-0.07}$      &      26.40$^{+0.03}_{-0.03}$   &     26.26$^{+0.08}_{-0.08}$ \\
\hline
3.0$\pm{0.24}$$^d$     &      25.58$^{+0.31}_{-0.17}$     &       &     26.22 \\
\enddata
\tablenotetext{a}{units: erg s$^{-1}$ Hz$^{-1}$ Mpc$^{-3}$, flat-lambda cosmology with $H_0 = 70$ km s$^{-1}$ Mpc$^{-1}$ ($\Omega_M = 0.3, \Omega_\Lambda = 0.7$)}
\tablenotetext{b}{$L_{min}=0.2L_{\star} (z=3)^d;  M_{min} = -19.32$}
\tablenotetext{c}{data from Wyder et al. (2004)}
\tablenotetext{d}{data from Steidel et al. (1999)}
\tablenotetext{e}{fixed faint end slope $\alpha = -1.6$}

\label{tab:bcseds}
\end{deluxetable}

\begin{figure}
\plotone{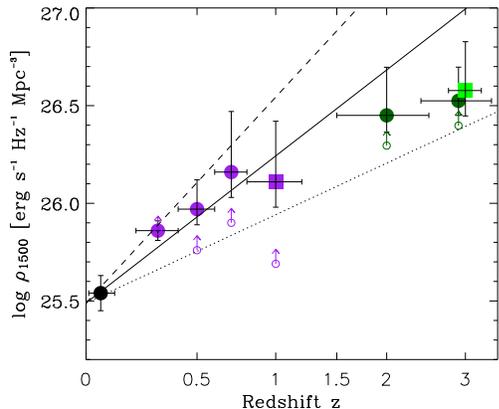}
\caption{FUV  luminosity density vs. redshift.  Filled circles indicate LF fit to full sample with unconstrained slope $\alpha$.  Filled squares denote LF fit with fixed $\alpha =-1.6$.  Purple (GALEX) and dark green (HDF) symbols are from this work. Black dot is taken from local LF (Wyder et al. 2004) and green square from Steidel et al. (1999).  Open circles denote $\rho_{1500}$ determined using V$_{max}$.  Errors do not include cosmic variance.  Lines indicate $(1+z)^n$ evolution.  Dotted, solid and dashed lines correspond to $n=1.5, 2.5, 3.5$ respectively.   
\label{fig1}}
\end{figure}

\begin{figure}
\plotone{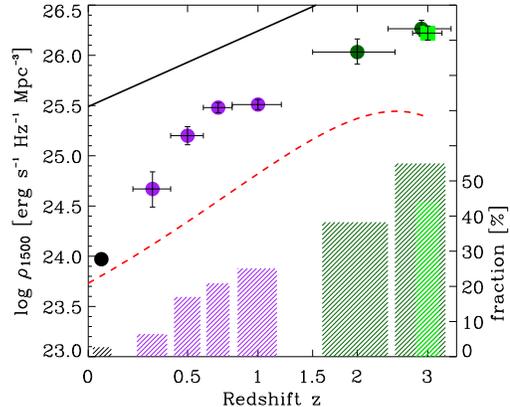} 
\caption{FUV luminosity density of ultraviolet luminous galaxies (UVLGs) vs. redshift and comparison w/ QSO luminosity density. Filled circles from $\rho_{1500,UVLG}$ integrated from $\infty$ down to 0.2L$_{\star,z=3}$ or  M$_{FUV}$ -19.32.   Colors same as Figure 1.  Vertical hatched bars indicate fraction of luminosity emitted by galaxies brighter than 0.2$L_{\star,z=3}$.  Red dashed line shows QSO FUV LD using values from Boyle et al. (2000) and Madau et al. (1999).  Solid line same as Figure 1.  
\label{fig2}}
\end{figure}

\section{Luminosity density}

We calculated the FUV luminosity density $\rho_{1500}(z)$ from the
GALEX-VVDS sample in four redshift bins ($<z>=0.3, 0.5, 0.7, 1.0$), and also determined values
for the HDF sample ($<z>=2.0, 2.9$).  Results are shown in Table 1 and plotted in Figure 1.  We
chose to calculate $\rho_{1500}(z)$ in several ways.  First we summed
$\phi(L) LdL$ using the LF obtained from the V$_{max}$ method.
Because luminosity bins with no detections do not contribute
we consider this a lower limit on $\rho_{1500}(z)$.  We also
calculated a ``total'' luminosity density by integrating Schechter
function fits to the LF
using the formula: $$\rho=\int_{L_{min}}^{\infty} dLL\phi(L) $$ with
$L_{min} = 0$ ($\rho= \phi_*L_*\Gamma(\alpha+2)$). Although this
quantity is strongly dependent on uncertainties in the faint end slope
($\alpha$), it allows direct comparison with other measurements of
$\rho_{1500}(z)$ and the star formation rate density, $\dot{\rho_\star}$.  Fits and errors were determined using the ALF tool (Ilbert et al., 2004)
with error bars based on the extreme values of the LD calculated at each point on the $\alpha$-M$_\star$ 1$\sigma$ error contour.
For the z=1.0 bin, our best
Schechter function fit yielded large errrors for the slope
($\alpha$=-1.63$^{+0.45}_{-0.43}$).  For this bin we fixed the faint end slope at $\alpha$ to -1.6, adopting the value used in high-z studies (e.g. Steidel et al. 1999) and
consistent within errors with our own values at lower and higher z.
Total $\rho_{1500}(z)$ shows significant $\sim(1+z)^{2.5}$
evolution out to z$\sim$1, with evidence for a shallow continued rise out to
z=3.  This evolution is discussed further in the next section.  
Two points are worth noting regarding the comparison of LD at different redshifts.  First, as demonstrated  in (Paper I) and discussed below, the galaxy population that contributes most of the LD varies (vs. color, luminosity) with redshift. Secondly, while most of the sample is UV-selected, the Steidel et al. (1999) LBG galaxies were color-selected and the z=3 LD value may be missing some fraction of the UV light.  The similarity between the z=2.9 and z=3 data points suggests that the missing fraction is small.

We explore the contribution to the luminosity density from UV
luminous galaxies (UVLGs) by measuring the luminosity density  from galaxies
with $L>L_{min}$.  To facilitate comparison with high-z
studies, we set $L_{min} = 0.2L_{\star, z=3}$ ($M_{min} = -19.32$) from Steidel et al. (1999), also adopted by Giavalisco et al.
(2004) for their work.\footnote{This luminosity corresponds to 10$^{10.1}$ L$_{\odot}$, $\sim2/3$ the luminosity limit (10$^{10.3}$ L$_{\odot}$) adopted for UVLGs in Heckman et al (2004).}  These galaxies are observable in all redshift ranges and therefore there is no additional uncertainty related to extrapolation beyond the faintest observed magnitude.  Figure 2 highlights the dramatic evolution of
$\rho_{1500,UVLG}$, increasing by $\times$30 to z$\sim$1 or
(1+z)$^5$.  Furthermore we find that UVLGs are a major
contributor to $\rho_{1500}$ at z$\sim$1
with a fractional contribution, $\rho_{1500,UVLG}/\rho_{1500}$  of 25\%.  
We plot for comparison $\rho_{1500,QSO}$ using the functional form of the QSO LD evolution (in the B-band) from Boyle et al. (2000) and the QSO SED from Madau, Haardt, \&  Rees (1999) which has a shallower evolutionary slope (3.5) vs. UVLGs (5) for z$<$1.  

\begin{figure}
\plotone{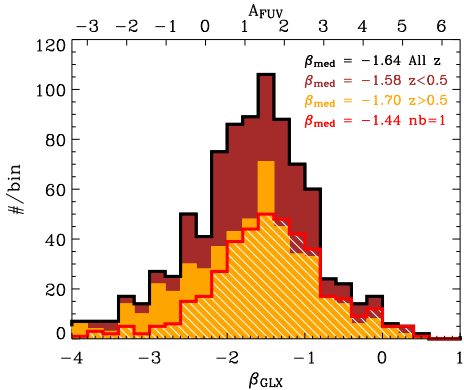}
\caption{Histogram of FUV slope $\beta_{GLX}$ for 888 galaxies with measurements in U-band ($black$) and split into two subsamples: $z<$0.5 ($brown$) and z$>$0.5 ($orange$).   ($Red/white$) Distribution of $\beta_{GLX}$ for ``isolated'' GALEX detections with only one optical counterpart within 4$^{\prime\prime}$ radius.  
\label{fig3}}
\end{figure}

\begin{figure}
\plotone{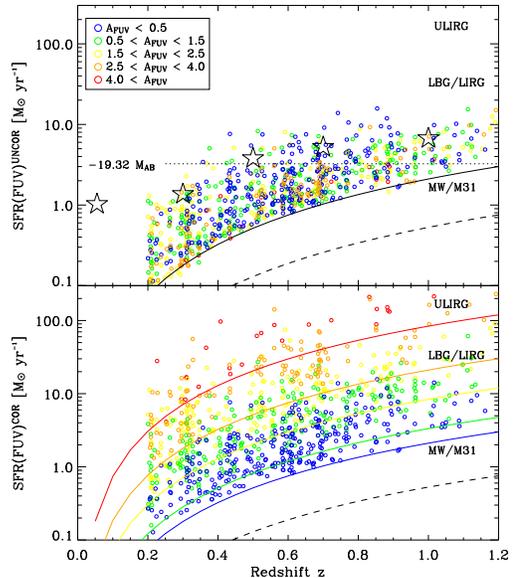}
\caption{Star formation rate of GALEX-VVDS galaxies vs. redshift using Kennicutt (1998) conversion.  Top: Uncorrected star formation rate for galaxies with varying A$_{FUV}$.  Solid line and dashed line correspond to NUV$<$24.5 (current sample) and NUV$<$26 (GALEX Ultra-deep survey) limits.  Stars show values for L$_{\star}$ from Paper I and Wyder et al. (2004).  Dotted line corresponds to $L_{min} = 0.2L_{\star,z=3}$ cut.  Bottom: attenuation-corrected star formation rates.  Symbols same as top.  Colored lines correspond to detection limits for NUV$<$24.5 at minimum attenuation level for each subsample.
\label{fig:sfr}}
\end{figure}

\section{Star formation rate density}

To determine intrinsic ultraviolet luminosities for the GALEX-VVDS sample, we apply the MHC99 dust attenuation formula:
$$A_{FUV} = 4.43+1.99 (\beta) = 4.49+1.97 (\beta_{GLX}) $$
where we use the definition of $\beta_{GLX}$, the FUV slope calculated using the rest-frame GALEX FUV and NUV bands, from Kong et al. (2004).  We only calculate $\beta_{GLX}$ for the subset of galaxies observed in U-band (888 galaxies).  Typical errors are $\sigma_{\beta}\sim0.4$.  Figure 3 shows the distribution of the k-corrected $\beta_{GLX}$.  The full sample has median $\beta_{GLX} = -1.64$, $FWHM(\beta)=1.4$ with little variation with redshift.  We find good agreement with measurements of $\beta$ at low-z ($<\beta>=-1.6$ for a FUV selected sample; Treyer et al. 2004) and high-z (Adelberger \& Steidel 2000).

Within our own sample we might have expected to see an increase of $\beta_{GLX}$ vs. z since high luminosity galaxies---which dominate the high-z bins---are expected to show significant attenuation.  Several effects could work against this trend.  We are detecting galaxies close to the NUV band confusion limit (beam/source$\sim$10 for NUV$<$25) and source blending could shift UV-optical colors and the slope {\it blueward}.   We performed tests which conservatively apportioned NUV flux among all potential optical counterparts and set a limit on the offset of the median $\Delta\beta_{GLX,blend}\le0.35$. This is consistent with the median $\beta_{GLX}=-1.44$ measured for ``isolated'' UV detections with only a single optical counterpart (see Figure 3).  (However, we can't neglect the possibility that some fraction with multiple counterparts are physical pairs which could show a different distribution of $\beta_{GLX}$).  We also note that the MHC99 $A_{FUV}$-$\beta$ relation was determined for starbursting galaxies (the bulk of our sample, see Paper I) but might overestimate the correction for normal star forming galaxies (Bell 2002; Kong et al. 2004) which are found in our lowest redshift bins.  For a conservative measurement of the average attenuation in our whole sample, we use the ``isolated'' subsample, and calculate a mean attenuation factor of $\times$7 (A$_{FUV}^{meas}=1.8$) where we have estimated and applied a bias correction to the mean ($\times0.7$) due to non-negligible $\sigma_{\beta}$.   We also adopt a 'minimum attenuation' A$_{FUV}^{min}$=1 which may be more representative of a full UV-selected population (Buat et al. 2004).

\begin{figure}
\plotone{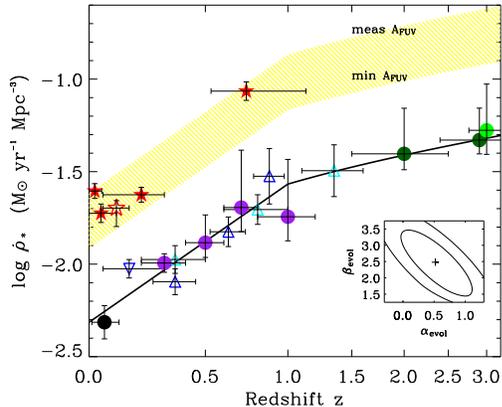}
\caption{Star formation rate density vs. z.  Filled circles from measurements at 1500\AA~(uncorrected for dust) same as in Figure 1. Blue comparison points are rest-frame UV measurements uncorrected for dust attenuation.  Inverted blue triangle from Sullivan et al. 2000.  Dark blue triangles from Lilly (1996).  Light blue triangles from Wilson et al. (2002) for $\alpha=-1.5$.  Solid line rises as (1+z)$^{2.5}$ for z$<$1 and then (1+z)$^{0.5}$ for z$>$1 based on chi-squared fit to our sample (see inset; 1$\sigma$ and 2$\sigma$ confidence contours shown).  Shaded region shows range corresponding to max/min dust-attenuation.  Filled red stars from dust-corrected H$\alpha$ measurements (with increasing redshift) from P\'erez-Gonz\'alez et al. (2003), Gronwall (1999), Tresse \& Maddox (1998), Tresse et al. (2002).  Open red star from SDSS (H$\alpha$/emission line) Brinchmann et al. 2004.  
\label{fig4}}
\end{figure}

The SFR was calculated for each galaxy using:
$$SFR (M_{\odot}~{\rm yr}^{-1}) = 1.4 \times 10^{-28} L_{FUV} \rm{(erg~s^{-1}~Hz^{-1})}  $$
from Kennicutt (1998).  In Figure 4 we plot the SFR derived for each galaxy using the  uncorrected and the dust-corrected FUV luminosities.  Our sample shows no dependence of dust attenuation with  SFR$^{uncor}$ and as a consequence we find higher attenuation in galaxies with high SFR$^{cor}$.  This paucity of low-attenuation galaxies with high SFR$^{cor}$ has been noted in previous studies (e.g. Wang \& Heckman 1996, Adelberger \& Steidel  2000).  Some of the observed effect may also be due to the scatter in A$_{FUV}$ discussed above (resulting in a tail of high A$_{FUV}$ galaxies) and/or limitations of the dust attenuation law.   We plot $\dot{\rho_\star}(z)$ (derived from $\rho_{1500}$ with no dust correction) in Figure 5.  Measurements from this paper were fit using the parametrization from Baldry et al. (2002)  ($\dot{\rho_\star}(z)\sim(1+z)^{\beta_{evol}}, z<1$ and $\dot{\rho_\star}(z)\sim(1+z)^{\alpha_{evol}}, z>1$).  We find a best-fit $\beta_{evol}=2.5\pm0.7$, $\alpha_{evol}=0.5\pm0.4$.  The 1$\sigma$ constraint on the ($\alpha_{evol}, \beta_{evol}$) pair is consistent with independent derivations using 2dF (Baldry et al. 2002), SDSS (Brinchmann et al. 2003) and other recent studies (c.f. Fig. 13 Baldry et al. 2002). 

Several uncorrected ($blue$) and dust-corrected ($red$) comparison measurements obtained using spectroscopic redshifts are shown in Figure 5.  Before determining $\dot{\rho_\star}$ we converted $\rho_{2000}$ (Sullivan et al. 2000, Lilly et al. 1996) and $\rho_{2500}$ (Wilson et al. 2002, $\alpha=-1.5$ data) to $\rho_{1500}$ using $\rho(\lambda)$ obtained from local $\rho_{1540}$ and $\rho_{2300}$  by Wyder et al. 2004 ($\sim\lambda^{0.9}$).  Wilson et al. (2002)  and Lilly et al. (1996) both show good agreement with our measured values despite the difference in  evolutionary slope obtained in the two studies ($\beta_{evol}\sim1.7\pm1$, $3.3\pm0.7$, respectively) .  The local luminosity density reported by Sullivan (2000) appears high, as noted in Wyder et al. (2004). 
Finally, we show a likely range of dust-corrected SFR densities, applying the average A$_{FUV}^{min}$, A$_{FUV}^{meas}$ to the best-fit parametrized $\dot{\rho_\star}(z)$.   Using the Kennicutt (1998) SFR conversion, we find that recent dust-corrected H$\alpha$ measurements fall within our attenuation-corrected range.  Although we have implicitly assumed no evolution in the dust correction, we emphasize that for UV flux-limited samples we might expect evolution in the average dust-attenuation correction vs. redshift and will explore this further in future work.

The FUV is
tracing a predominantly homogeneous population (star-forming and starbursting) making interpretation of integrated measures much more straightforward than at longer wavelengths (cf. Wolf et al. 2003).   We have shown that a significant population of UVLGs lies within easy reach ($0.6<z<1.2$).  We will compare these unique star-forming galaxies with their high-redshift LBG analogs (e.g. Shapley et al. 2003). In the near future our sample will expand by  $\times$5 in this field alone, and by more than $\times$100 using data from redshift surveys across the sky.  In some locations we will increase our depth to m$_{AB}\sim26$ as part of the Ultra-Deep Imaging Survey and probe down to 0.1L$_\star$  (see Figure 4) to better constrain the faint end of $\phi_{FUV}$. This will be supplemented by an even larger catalog ($>$10$^6$ objects) with photometric redshifts.  We will soon be able determine how SFR evolution depends on environment, morphology and spectral type and will examine our results within the context of cosmological simulations.  A major challenge lies in the understanding the role of dust obscuration, one which we will explore using recent, more sophisticated models (e.g. Kong et al. 2004) as the GALEX surveys continue.

\acknowledgments

GALEX (Galaxy Evolution Explorer) is a NASA Small Explorer, launched in April 2003.
We gratefully acknowledge NASA's support for construction, operation, and science analysis for the GALEX mission, developed in cooperation with the Centre National d'Etudes Spatiales of France and the Korean Ministry of Science and Technology.

Ê

Ê

Ê


%% copy editing.  Individual instruments can be provided in parentheses,
%% after the keyword, but they will not be verified.

%% Facilities: \facility{GALEX},\facility{VLT/VVDS}

\clearpage


\begin{thebibliography}{}

\bibitem[Adel(2000)]{ade00}Adelberger, K. L., \& Steidel, C. C. 2000, \apj, 544, 218
\bibitem[Arnouts(2002)]{arn02} Arnouts, S. et al. 2002, \mnras,   329, 355
\bibitem[Arnouts(2004)]{arn04} Arnouts, S. et al. 2004, \apjl,    this volume (Paper I)
\bibitem[Baldry(2002)]{bal02} Baldry, I. K. et al. 2002, \apj, 569, 582
\bibitem[Bell2002]{bell2002}Bell, E. 2002, \apj, 577, 150
\bibitem[Boyle(2000)]{boy00} Boyle, B. J., Shanks, T., Croom, S. M., Smith, R. J., Miller, L., Loaring, N., \& Heymans, C. 2000, \mnras, 317, 1014
\bibitem[Brinchmann(2003)]{bri03} Brinchmann, J.,  Charlot, S., White, S. D. M., Tremonti, C., Kauffmann, G., Heckman, T., \& Brinkmann, J. 2004, MNRAS in press
\bibitem[Buat(2004)]{bua04} Buat, V. et al. 2004, \apjl,    this volume
\bibitem[Cow(1999)]{cow99} Cowie, L. L., Songaila, A., Barger, A. J., 1999, \aj, 118, 603
\bibitem[Giav(2004)]{gia04} Giavalisco, M. et al. 2004, \apjl, 600, L103
\bibitem[Gronwall(1999)]{gro99} Gronwall, C., 1999, in Proceedings of the Conference ``After the Dark Ages: When Galaxies were Young'', ed. S. Holt and E. Smith, AIP, p. 335
\bibitem[Heckman(2004)]{hec04} Heckman, T. M. et al. 2004, \apjl,    this volume 
\bibitem[Hopkins(2004)]{hop04} Hopkins, A. M. 2004, \apj, in press
\bibitem[Ilbert(2004)]{Ilb04} Ilbert et al., 2004, AA, submitted
\bibitem[Kennicutt(1998)]{ken98} Kennicutt, R. C., Jr. 1998, \araa,    36, 189
\bibitem[Kong(2004)]{kon04} Kong, X., Charlot, S., Brinchmann, J., \& Fall, S. M. 2004, \mnras, 349, 769
\bibitem[Lilly(1996)] {lil96}Lilly, S. J., Le Fevre, O., Hammer, F., \& Crampton, D. 1996, \apjl, 460, L1
\bibitem[Madau(1999)]{mad99}Madau, P., Haardt, F., \& Rees, M. 1999, \apj, 514, 648
\bibitem[Madau(1998)]{mad98}Madau, P., Pozzetti, L., \& Dickinson, M. 1998, \apj, 498, 106
\bibitem[Martin(2004)]{mar04} Martin, D. C. et al, 2004, \apjl,  this volume
\bibitem[Meu(1999)]{meu99} Meurer, G. R., Heckman, T. M., \& Calzetti, D. 1999, \apj, 521, 64
\bibitem[Perez2003]{pg03}P\'erez-Gonz\'alez, P. G., Zamorano, J., Gallego, J., Arag\'on-Salamanca, A., \& Gil de Paz, A. 2003, \apj, 591, 827
\bibitem[Shapley(2003)]{sha03} Shapley, A. E., Steidel, C. C., Pettini, M., \& Adelberger, K. 2003, 588, 65
\bibitem[Steidel(1999)]{ste99} Steidel, C. C., Adelberger, K. L., Giavalisco, M., Dickinson, M., \& Pettini, M. 1999, \apj,    519, 1
\bibitem[Sullivan(2000)]{sul00}Sullivan, M., Treyer, M. A., Ellis, R. S., Bridges, T. J., Milliard, B., \& Donas, J. 2000, \mnras, 312, 442
\bibitem[Tresse(1998)]{tre98} Tresse, L., \& Maddox, S. J. 1998, \apj, 495, 691
\bibitem[Tresse(2002)]{tre02} Tresse, L., \& Maddox, S. J., Le Fevre, O., \& Cuby, J.-G. 2002, \mnras, 337, 369
\bibitem[Treyer(2004)]{tre04} Treyer, M. A. et al. 2004, \apjl,    this volume
\bibitem[Wang(1996)]{wh96} Wang, B. \& Heckman, T. M. 1996, \apj, 457, 645
\bibitem[Wilson(2002)]{wil02} Wilson, G., Cowie, L. L., Barger, A. J., \& Burke, D. J. 2002, \aj, 124, 1258
\bibitem[Wolf(2003)]{wol03} Wolf, C., Meisenheimer, K., Rix, H.-W., Borch, A., Dye, S., \& Kleinheinrich, M. 2003, \aap,    401, 73
\bibitem[Wyder(2004)]{wyd04} Wyder, T. et al, 2004, \apjl, this volume

Ê
\end{thebibliography}
\end{document}